%% file: main.tex
\documentclass{eptcs}
\providecommand{\event}{ICE 2014} 

\usepackage{cite}
\usepackage{subfig}
\usepackage{sung}
\usepackage{tikz-reo}
\usepackage{relsize}

\input{cmds}

\newcommand{\COMPENSATE}{\vspace{-10pt}}

\title{Toward Sequentializing Overparallelized Protocol Code}
\author{Sung-Shik T.Q. Jongmans \qquad\qquad Farhad Arbab
\institute{Centrum Wiskunde \& Informatica \\ Amsterdam, Netherlands}
\email{[jongmans,farhad]@cwi.nl}
}
\def\titlerunning{Toward Sequentializing Overparallelized Protocol Code}
\def\authorrunning{S.-S.T.Q. Jongmans \& F. Arbab}

\begin{document}

\maketitle

\begin{abstract}
	In our ongoing work, we use constraint automata to compile protocol specifications expressed as Reo connectors into efficient executable code, e.g., in C.
	We have by now studied this automata based compilation approach rather well, and have devised effective solutions to some of its problems. 
	Because our approach is based on constraint automata, the approach, its problems, and our solutions are in fact useful and relevant well beyond the specific case of compiling Reo. 
	In this short paper, we identify and analyze two such rather unexpected problems.
\end{abstract}

%
\section*{Introduction}
\label{sect:intr}

A promising application domain for coordination languages is programming protocols among threads in multicore programs: coordination languages typically provide high-level constructs and abstractions that more easily compose into correct---with respect to a programmer's intentions---protocol specifications than do low-level synchronization constructs provided by conventional languages (e.g., locks, semaphores).
In fact, not only do coordination languages simplify programming protocols, but their high-level constructs and abstractions also leave more room for compilers to perform novel optimizations in mapping protocol specifications to lower-level instructions that implement them.
A crucial step toward adoption of coordination languages for multicore programming is the development of such compilers: programmers need tools to generate efficient code from high-level protocol specifications.

\begin{figure}
	\COMPENSATE
	\newcommand{\HEIGHT}{105pt}
	\newcommand{\SCALE}{.4}
	
	\hfil
	\subfloat[\reo{AsyncMerger}]{\label{fig:conn:asyncmerger}%
		\vbox to \HEIGHT {%
			\vfil \hbox {\includegraphics[scale=\SCALE]{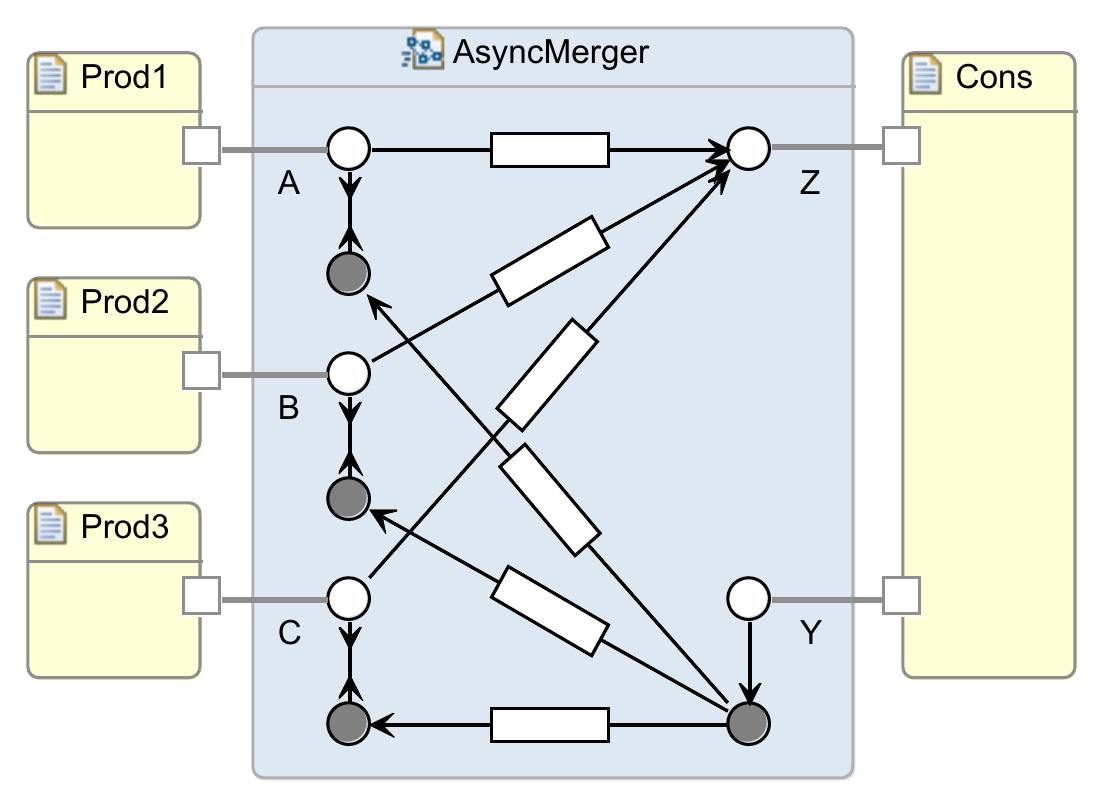}} \vfil
		}
	}
	\hfil
	\subfloat[\reo{Alternator}]{\label{fig:conn:alternator}%
		\vbox to \HEIGHT {%
			\vfil \hbox {\includegraphics[scale=\SCALE]{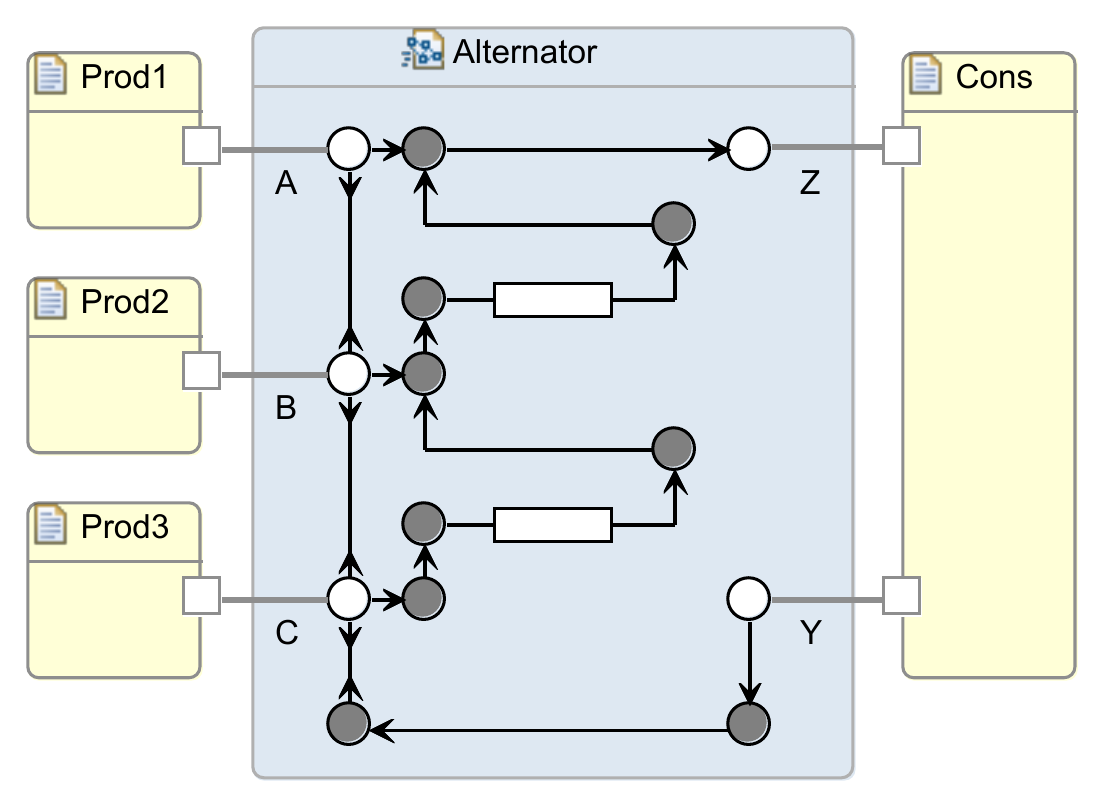}} \vfil
		}
	}
	\hfil
	\subfloat[Synchr. region of \reo{Alternator}]{\label{fig:conn:alternator-sr}%
		\vbox to \HEIGHT {%
			\vfil \hbox {\includegraphics[scale=\SCALE]{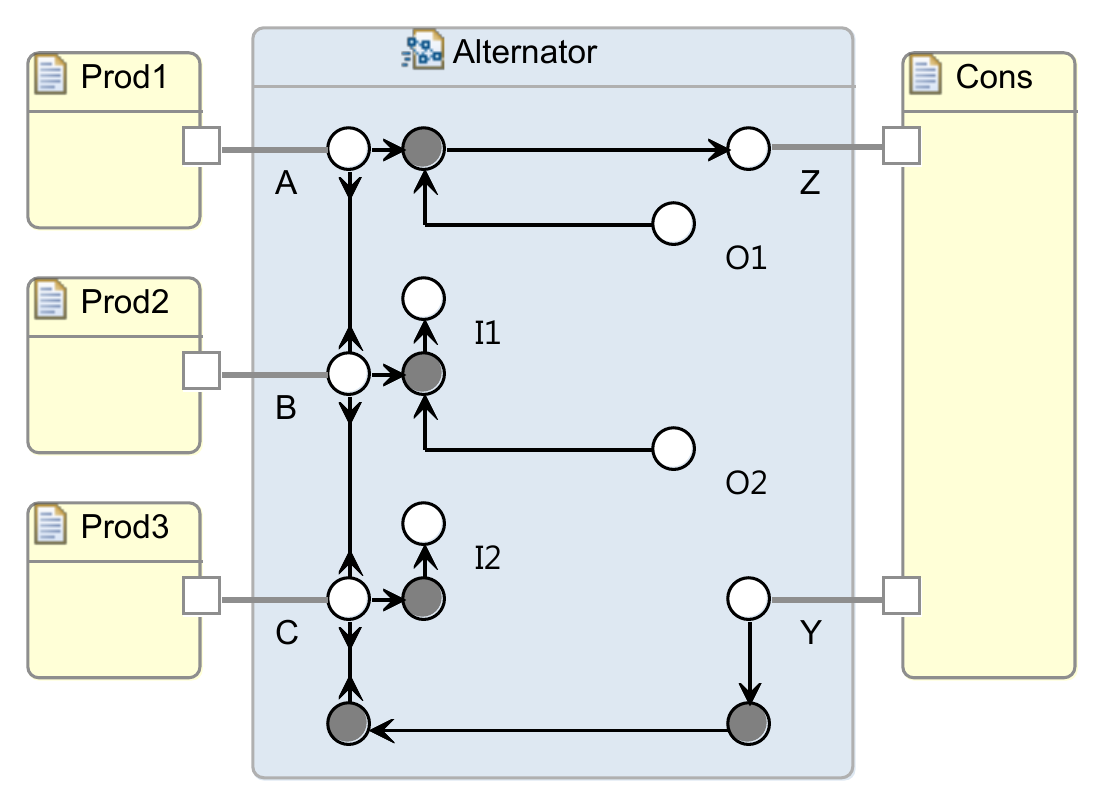}} \vfil
		}
	}
	\hfil
	
	\caption{Example connectors}
	\label{fig:conn}
\end{figure}

In ongoing work, we develop compiler technology for the graphical coordination language Reo~\cite{Arb11}.
Reo facilitates compositional construction of protocol specifications manifested as \emph{connectors}: channel-based mediums through which threads can communicate with each other.
Figure~\ref{fig:conn} shows some example connectors, each linked to four \emph{computation threads}, in their usual graphical syntax.
Briefly, a connector consists of one or more \emph{channels}, through which data items flow, and a number of \emph{nodes}, on which channel ends coincide.
In Figure~\ref{fig:conn}, we distinguish the \emph{boundary nodes} of a connector (to which computation threads are linked) from its \emph{internal nodes} (used only for internally routing data) by shading the internal nodes.
The connectors in Figure~\ref{fig:conn} contain three different channel classes, including standard synchronous channels (normal arrows) and asynchronous channels with a buffer of capacity 1 (arrows decorated with a white rectangle, which represents a buffer).
Through connector \emph{composition} (the act of gluing connectors together on their shared nodes), programmers can construct arbitrarily complex connectors.
As Reo supports both synchronous and asynchronous channels, connector composition enables mixing synchronous and asynchronous communication within the same protocol specification.

Figure~\ref{fig:conn:asyncmerger} shows a connector, \reo{AsyncMerger}, for a protocol among $k = 3$ producers and one consumer.
We compared the code generated by our Reo-to-C compiler~\cite{JHA13} with hand-crafted code written by a competent C programmer using Pthreads, investigating the time required for communicating a data item from a producer to the consumer as a function of the number of producers $4 \leq k \leq 512$.

The results looked excellent: the code generated by our compiler outperforms the hand-crafted code and scales well~\cite{JHA14}.
Encouraged by this outcome, we expected to reproduce these results for the producers--consumer protocol specified by the \reo{Alternator} connector in Figure~\ref{fig:conn:alternator}.%
\footnote{%
	In the \reo{AsyncMerger} protocol. the consumer receives productions in arbitrary order. In contrast, in the \reo{Alternator} protocol, the consumer receives data ``from top to bottom'' (and to achieve this, the producers collectively synchronize before sending).
}
The results disappointed us: for small $k$, the code of \reo{Alternator} runs significantly slower than that of \reo{AsyncMerger}, while for large $k$, the compiler times out (i.e., after five minutes, we manually aborted the compilation process).

In this short paper, we identify two ``unexpected'' problems of our current compilation approach (which manifest in \reo{Alternator}): exponential explosion at compile-time and overparallelization at run-time.
These problems are in fact unfortunate side effects of another optimization step in our compilation process that we thought we had well studied.
After an analysis, we propose a first solution that works in some---but not all---problematic cases; we leave a comprehensive solution for future work and consider the identification and analysis of the two problems the main contribution of this short paper.

%
\section*{Problem Analysis and a First Solution}

\begin{figure}
	\COMPENSATE
	\newcommand{\HEIGHT}{62pt}
	
	\hfil
	\subfloat[\reo{Alternator}]{\label{fig:ca:alternator}%
		\vbox to \HEIGHT {%
			\vfil \hbox {\scalebox{1}{\begin{tikzpicture}[baseline, node distance=1.25cm]
				\node[state, pin={[pin edge={black, thick, <-}]above left:}] (Q1) {};
				\node[state, right of=Q1] (Q2) {};
				\node[state, right of=Q2] (Q3) {};
				
				\path[->] (Q1) edge [trans, out=30, in=150] node [above] {\scriptsize $\set{\reo{A} \, \reo{B} \, \reo{C} \, \reo{Y} \, \reo{Z}}$} (Q3);
				\path[->] (Q3) edge [trans, out=-150, in=-30] node [below] {\scriptsize $\set{\reo{Z}}$} (Q2);
				\path[->] (Q2) edge [trans, out=-150, in=-30] node [below] {\scriptsize $\set{\reo{Z}}$} (Q1);
			\end{tikzpicture}}} \vfil
		}
	}
	\hfil
	\subfloat[Synchronous region of \reo{Alternator}]{\label{fig:ca:alternator-sr}%
		\vbox to \HEIGHT {%
			\vfil \hbox {\quad \scalebox{1}{\tikzautonee{$\set{\reo{A} \, \reo{B} \, \reo{C} \, \reo{Y} \, \reo{Z} \, \reo{I1} \, \reo{I2}}$}{$\set{\reo{I1} \, \reo{O2}}$}{$\set{\reo{Z} \, \reo{O1} \, \reo{I1} \, \reo{O2}}$}{$\set{\reo{Z} \, \reo{O1}}$}} \quad} \vfil
		}
	}
	\hfil
	
	\caption{Example constraint automata (irrelevant details of transition labels omitted)}
	\label{fig:ca}
\end{figure}
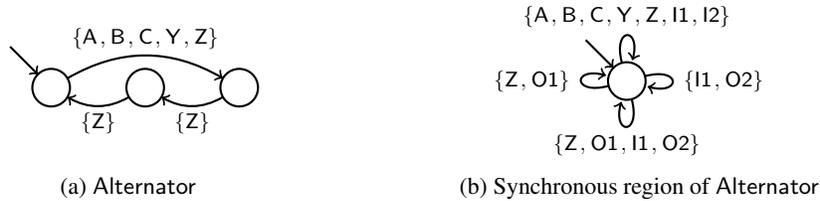

Our Reo-to-C compiler generates code for Reo connectors based on their \emph{constraint automaton} (\CA) semantics~\cite{BSAR06}.
Constraint automata are a general formalism for modeling systems, better suited for data-aware modeling of Reo connectors and, in particular, their composition (which supports multiparty and transitive synchronization) than classical automata or traditional process calculi.
Figure~\ref{fig:ca:alternator} shows examples.
For Reo, a \CA specifies \emph{when} during execution of a connector \emph{which} data items flow \emph{where}.
Structurally, every \CA consists of finite sets of states and transitions.
A \emph{product operator} on \CA, which preserves \CA-bisimilarity~\cite{BSAR06}, models connector composition: to obtain the ``big'' \CA for a whole connector, one can compute the product of the ``small'' \CA for its constituent nodes and channels.
Afterward, one can abstract away internal nodes with a \emph{hide operator} on \CA~\cite{BSAR06}, which---importantly---also eliminates silent transitions involving only internal nodes in a semantics-preserving way.

Although motivated by our work on Reo, our compiler really operates primarily at the level of Reo's \CA semantics.
In that sense, ``Reo-to-C compiler'' is a misnomer.
A better name would be ``\CA-to-C compiler'': we use Reo, with its graphical, channel-based abstractions, just as \emph{a}---not \emph{the}---programmer-friendly syntax for exposing \CA-based protocol programming.
Different syntax alternatives for CA may work equally well or yield perhaps even more user-friendly languages.
For instance, we know how to translate \UML sequence/activity diagrams and \BPMN to \CA~\cite{AKM08,CKA10,MAB11}.
Another interesting potential syntax are algebras of Bliudze and Sifakis~\cite{BS10}, originally developed in the context of \BIP~\cite{BBS06}, which have a straightforward interpretation in terms of \CA.
Due to their generality, \CA can thus serve as an intermediate language (transparent to programmers) for compiling specifications in many different languages and models of concurrency by reusing the core of our compiler.
This makes the development of this compiler and its optimizations relevant beyond Reo.

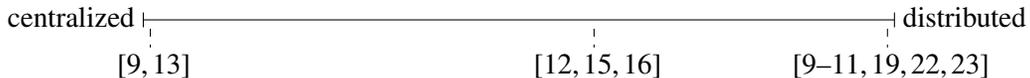
\begin{figure}
	\centering
	\begin{tikzpicture}
		\coordinate [pin={[inner sep=0pt, pin distance=.1cm, pin edge={white}]left:\begin{tabular}{@{}c@{}}centralized\end{tabular}}] (Cent);
		\coordinate [right of=Cent, node distance=10cm, pin={[inner sep=0pt, pin distance=.1cm, pin edge={white}]right:\begin{tabular}{@{}c@{}}distributed\end{tabular}}] (Dist);
		
		\draw [|-|] (Cent) to node {} (Dist);
		
		\node [right of=Cent, node distance=.1cm, inner sep=0cm, outer sep=.1cm, pin={[inner sep=2pt, pin distance=.25cm, pin edge={black, dashed}]below:\cite{CCA07,JA13a}}] (A) {};
		\node [left of=Dist, node distance=.1cm, inner sep=0cm, outer sep=.1cm, pin={[inner sep=2pt, pin distance=.25cm, pin edge={black, dashed}]below:\cite{CCA07,CP12,CPLA11,Pro11,PCVA11,PCVA12}}] (B) {};
		\node [right of=Cent, node distance=6cm, inner sep=0cm, outer sep=.1cm, pin={[inner sep=2pt, pin distance=.25cm, pin edge={black, dashed}]below:\cite{JA13b,JHA13,JSA14}}] (C) {};
		
	\end{tikzpicture}
	
	\caption{Connector implementation spectrum}
	\label{fig:spectrum}
\end{figure}

Two opposite \CA-based approaches to implementing a connector \reo{Conn} exist.
In the \emph{distributed approach}, the compiler first finds a small \CA for every channel and every node that \reo{Conn} consists of and afterward generates a piece of sequential code for each of those small \CA.
At run-time, every piece of sequential code has its own thread, henceforth referred to as \emph{protocol threads}, and a distributed algorithm among those threads ensures their proper synchronization.
In the \emph{centralized approach}, after finding a collection of small \CA, the compiler forms the product of all those \CA to get a big \CA for \reo{Conn}, abstracts away all internal nodes, and finally generates one piece of sequential code for that big \CA.
For \CA-based implementations, these two approaches constitute the two ends of the \emph{connector implementation spectrum} in Figure~\ref{fig:spectrum}: the further we get to the right end of the spectrum, the more parallelism a connector implementation exhibits.
(For completeness, Figure~\ref{fig:spectrum} contains also references to Reo connector implementation approaches based on other formalisms---in particular, \emph{connector coloring} and \emph{coordination constraints}~\cite{CCA07,CP12,CPLA11,Pro11,PC13a,PC13b,PCVA11,PCVA12}---which work not exactly the same as just described for \CA.)

Neither the distributed approach nor the centralized approach is satisfactory.
For instance, the distributed approach suffers from high latency at run-time (because the distributed algorithm required for synchronizing the parallel protocol threads is expensive).
The centralized approach, in contrast, achieves low latency, but it suffers from state space explosion at compile-time (because a big \CA for a whole connector may have a number of states exponential in the number of its constituent channels) and oversequentialization at run-time (because simulating a big \CA with one thread serializes transitions that could have fired in parallel).
To solve these problems (i.e., strike a balance between run-time latency and parallelism), we extensively studied a \emph{middle ground approach} roughly in the center of the connector implementation spectrum.
In this approach, the compiler splits a connector into $m_1$ \emph{asynchronous regions} of purely asynchronous communication (e.g., each of the buffered channels in Figure~\ref{fig:conn}) and $m_2$ \emph{synchronous regions} of synchronous communication.%
\footnote{%
	Splitting into regions occurs at the level of small \CA, without knowledge of the input connector~\cite{JA13b,JSA14}.
}
The compiler subsequently forms products on a per-region basis, resulting in $m_1 + m_2$ ``medium'' \CA, and generates a piece of sequential code for each of them.
At run-time, every generated piece of code has its own thread, as in the distributed approach, but the distributed algorithm required for synchronizing those protocol threads has substantially lower costs.
Moreover, the middle ground approach mitigates state space explosion and oversequentialization.
For these advantages, we moved our compiler from the centralized approach to the middle ground aproach.

Unfortunately and unexpectedly, although the middle ground approach works well for \reo{AsyncMerger}, it fails for \reo{Alternator}.
We analyze why, as follows.
First, Figure~\ref{fig:conn:alternator-sr} shows the single synchronous region of \reo{Alternator}.
Because nodes \reo{I1}, \reo{I2}, \reo{O1}, and \reo{O2} lie on the boundary of this region, the compiler cannot abstract those nodes away.
Next, Figure~\ref{fig:ca:alternator-sr} shows the medium \CA for this region.
Its $\set{\reo{A} \, \reo{B} \, \reo{C} \, \reo{Y} \, \reo{Z} \, \reo{I1} \, \reo{I2}}$-transition and its $\set{\reo{Z} \, \reo{O1}}$-transition correspond to the $\set{\reo{A} \, \reo{B} \, \reo{C} \, \reo{Y} \, \reo{Z}}$-transition and the two $\set{\reo{Z}}$-transitions of the big \CA in Figure~\ref{fig:ca:alternator}.
The $\set{\reo{I1} \, \reo{O2}}$-transition of the medium \CA models an internal execution step---abstracted away in the big \CA---in which a data item flows from the bottom buffer into the top buffer.
Finally, the $\set{\reo{Z} \, \reo{O1} \, \reo{I1} \, \reo{O2}}$-transition of the medium \CA models an execution step in which its $\set{\reo{Z} \, \reo{O1}}$-transition and its $\set{\reo{I1} \, \reo{O2}}$-transition fire simultaneously by true concurrency.

Now, imagine a generalization of \reo{Alternator} from three producers to $k$ producers (by replicating parts of \reo{Alternator} in Figure~\ref{fig:conn:alternator} in the obvious way).
Such a connector has $k - 1$ buffers.
Consequently, the medium \CA for its single synchronous region has $k - 2$ transitions (among others), each of which models an internal execution steps where a data item flows from one buffer to the buffer directly above it.
Because any subset of those transitions may fire simultaneously by true concurrency, the medium \CA has roughly $2^{k-2}$ transitions.
The medium \CA for \reo{Alternator} with 512 producers consequently has over $10^{153}$ transitions---approximately $10^{73}$ times the estimated number of hydrogen atoms in the observable universe---such that merely representing this \CA in memory is already problematic (let alone compositionally computing it).
Thus, transition relation explosion at compile-time is a serious problem.

Now, suppose that we manage to successfully compile \reo{Alternator} for a sufficiently small number of $\ell$ producers.
At run-time, we have $\ell$ parallel protocol threads: one for \reo{Alternator}'s synchronous region and one for each of its $\ell - 1$ asynchronous regions.
But despite this parallel implementation, the big \CA of \reo{Alternator} in Figure~\ref{fig:ca:alternator} (for $\ell = 3$) implies that \reo{Alternator} in fact behaves sequentially.
In other words, we use parallelism---and incur the overhead that parallelism involves---to implement intrinsically sequential behavior.
Thus, overparallelization at run-time is another serious problem.

Interestingly, the centralized approach, which our compiler used to apply, does not suffer from transition relation explosion or overparallelization for a number of reasons.
First, overparallelization is trivially not a problem, because the centralized approach involves only one sequential protocol thread.

The second reason relates to the fact that enabledness of transitions in \reo{Alternator}'s synchronous region depends on the (non)emptiness of the buffers in its $k$ asynchronous regions: many transitions are in fact permanently disabled.
For instance, every ``true-concurrency-transition'' composed of $3 \leq x \leq k - 1$ transitions labeled with $\set{\reo{I}i \, \reo{O}i+1}$ (for some $i$), where data items flow upward through $x$ consecutive buffers, never fires: by Reo's semantics, the $x-2$ middle buffers cannot become empty and full again in the same transition, which would happen if this true-concurrency-transition were to fire.
A compiler can eliminate such permanently disabled transitions---and thereby mitigate transition relation explosion---by forming the product of all medium \CA for \reo{Alternator}'s synchronous and asynchronous regions (in a particular order), effectively computing one big \CA.
Exactly this happens in the centralized approach.

The third reason relates to abstraction of internal nodes and transitions.
In the middle ground approach, nodes shared between different regions do not count as internal nodes; they are boundary nodes and the compiler cannot abstract them away.
In contrast, in the centralized approach, all those boundary nodes between regions become internal nodes, which the compiler can abstract away.
Consequently, the compiler can eliminate more silent transitions involving only internal nodes---and thereby further mitigate transition relation explosion---by applying the hide operator.

Having moved our compiler from the centralized approach to the middle ground approach to avoid state space explosion and oversequentialization, now, we must find solutions for the unfortunate side effects of this move: transition relation explosion and overparallelization.

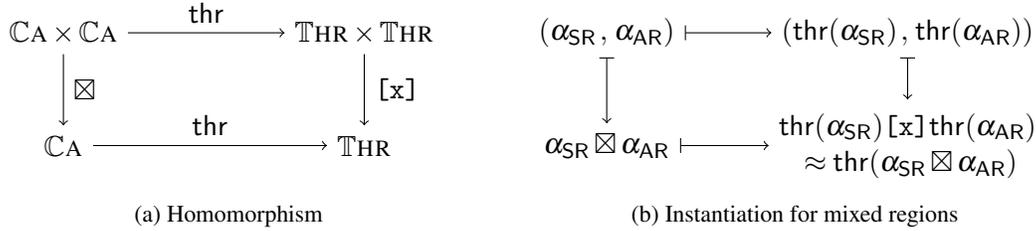
\begin{figure}
	\COMPENSATE
	\newcommand{\HEIGHT}{75pt}
	
	\hfil
	\subfloat[Homomorphism]{%
		\vbox to \HEIGHT {%
			\hbox {%
				\begin{tikzpicture}[node distance=4cm]
					\node[] (TL) {$\cauniv \times \cauniv$};
					\node[right of=TL] (TR) {$\thruniv \times \thruniv$};
					\node[below of=TL, node distance=1.5cm] (BL) {$\cauniv$};
					\node[below of=TR, node distance=1.5cm] (BR) {$\thruniv$};
					
					\draw[->] (TL) to node [above] {$\thrfun$} (TR);
					\draw[->] (TL) to node [right] {$\prodx$} (BL);
					\draw[->] (BL) to node [above] {$\thrfun$} (BR);
					\draw[->] (TR) to node [right] {$\prodxtt$} (BR);
				\end{tikzpicture}
			} \vfil
		}
	}
	\hfil
	\subfloat[Instantiation for mixed regions]{%
		\vbox to \HEIGHT {%
			\hbox {%
				\begin{tikzpicture}[node distance=4cm]
					\node[] (TL) {$\tpl{\alpha_\reo{SR} \, \alpha_\reo{AR}}$};
					\node[right of=TL] (TR) {$\tpl{\thrfun(\alpha_\reo{SR}) \, \thrfun(\alpha_\reo{AR})}$};
					\node[below of=TL, node distance=1.5cm] (BL) {$\alpha_\reo{SR} \prodx \alpha_\reo{AR}$};
					\node[below of=TR, node distance=1.5cm] (BR) {$\begin{array}{@{} c @{}} \thrfun(\alpha_\reo{SR}) \prodxtt \thrfun(\alpha_\reo{AR}) \\ {} \approx \thrfun(\alpha_\reo{SR} \prodx \alpha_\reo{AR}) \end{array}$};
					
					\draw[|->] (TL) to node [above] {\color{white} $\thrfun$} (TR);
					\draw[|->] (TL) to node [right] {\color{white} $\prodx$} (BL);
					\draw[|->] (BL) to node [above] {\color{white} $\thrfun$} (BR);
					\draw[|->] (TR) to node [right] {\color{white} $\prodxtt$} (BR);
				\end{tikzpicture}
			} \vfil
		}
	}
	\hfil
	
	\caption{Justification of mixed regions, where $\cauniv$ denotes the set of all \CA, $\thruniv$ denotes the set of all protocol threads, $\thrfun$ denotes a translation from \CA to protocol threads (i.e., actual code generation), $\prodx$ denotes the product operator on \CA, $\prodxtt$ denotes parallel composition of protocol threads synchronized by a distributed algorithm~\cite{JA13b,JSA14}, and $\approx$ denotes observational equivalence of protocol threads.}
	\label{fig:cat}
\end{figure}

Our first solution is to, at compile-time, merge every asynchronous region \reo{AR} that shares nodes with only one synchronous region \reo{SR} (i.e., \reo{AR} is neither connected to another region nor linked to a computation thread) into \reo{SR}.
Doing so results in a \emph{mixed region}.
Computation of mixed regions is semantics-preserving by the associativity and commutativity of the product operator on \CA~\cite{BSAR06}: if $\alpha_\reo{SR}$, $\alpha_\reo{AR}$, and $\alpha_\text{other}$ denote the \CA for \reo{SR}, \reo{AR}, and the other regions, the compiler can always change the bracketing of a product term over those \CA to a form in which $\alpha_\reo{SR}$ and $\alpha_\reo{AR}$ are the operands of the same product operator.
The compiler can subsequently decide either to actually form that product (thus computing the \CA of a mixed region) or leave $\alpha_\reo{SR}$ and $\alpha_\reo{AR}$ as separate \CA.
In the former case, at run-time, the protocol thread for the resulting product participates as one entity in the distributed algorithm for synchronizing protocol threads; in the latter case, both the protocol thread for $\alpha_\reo{SR}$ and the protocol thread for $\alpha_\reo{AR}$ participate in this algorithm.
Semantically, these implementations are indistinguishable.
More formally, the diagram in Figure~\ref{fig:cat} commutes.

Intuitively, forming mixed regions mitigates transition relation explosion at compile-time because (i) the compiler essentially computes a bigger product (which may eliminate permanently disabled transitions) and (ii) the compiler can abstract away more internal nodes (which may eliminate more silent transitions involving only internal nodes), namely all those shared between \reo{SR} and \reo{AR}.
Overparallelization at run-time is mitigated because every asynchronous region connected only to \reo{SR} \emph{must} interact with \reo{SR} in each of its transitions; it can never fire a transition independently of \reo{SR}.
Running such an asynchronous region in its own protocol thread would therefore never result in useful parallelism.

If we apply this first solution to \reo{Alternator}, the compiler merges all asynchronous regions into \reo{Alternator}'s single synchronous region.
This results in a single mixed region spanning the whole connector.
In this case, thus, the compiler reduces the middle ground approach back to centralized approach.

Although formulated generally in terms of regions, we know of cases of overparallelization that our first solution fails to mitigate.
For instance, although the \reo{Sequencer} connector has intrinsically sequential behavior~\cite{Arb11}, each of its asynchronous regions has connections to two---not one---synchronous regions.
We are thinking of generalizing our first solution to capture also this and similar cases, although we are not convinced yet that such a generalization exists; perhaps we need a rather different kind of rule.

\begin{figure}
	\COMPENSATE
	\newcommand{\HEIGHT}{151pt}
	\newcommand{\SCALE}{.4}
	
	\hfil
	\subfloat[\reo{Sync1}]{\label{fig:sync:1}%
		\vbox to \HEIGHT {%
			\vfil \hbox {\includegraphics[scale=\SCALE]{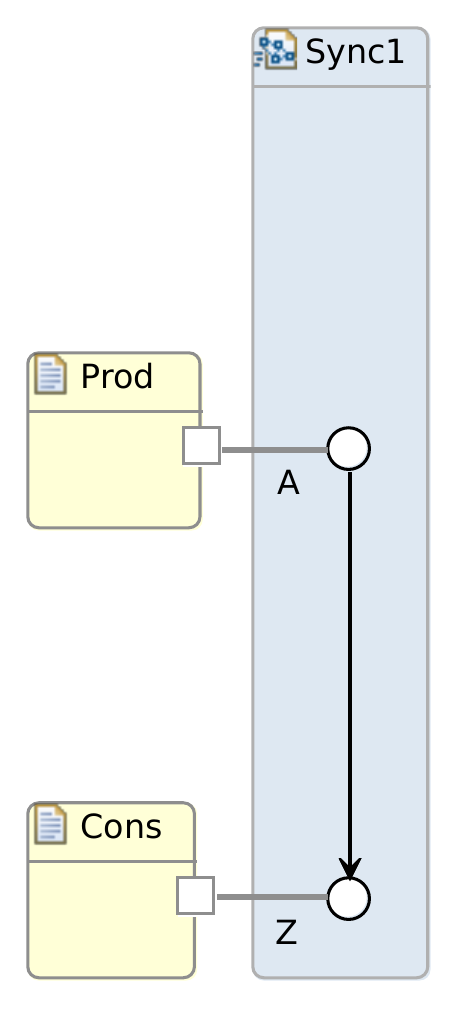}} \vfil
		}
	}
	\hfil
	\subfloat[\reo{Sync2}]{\label{fig:sync:2}%
		\vbox to \HEIGHT {%
			\vfil \hbox {\includegraphics[scale=\SCALE]{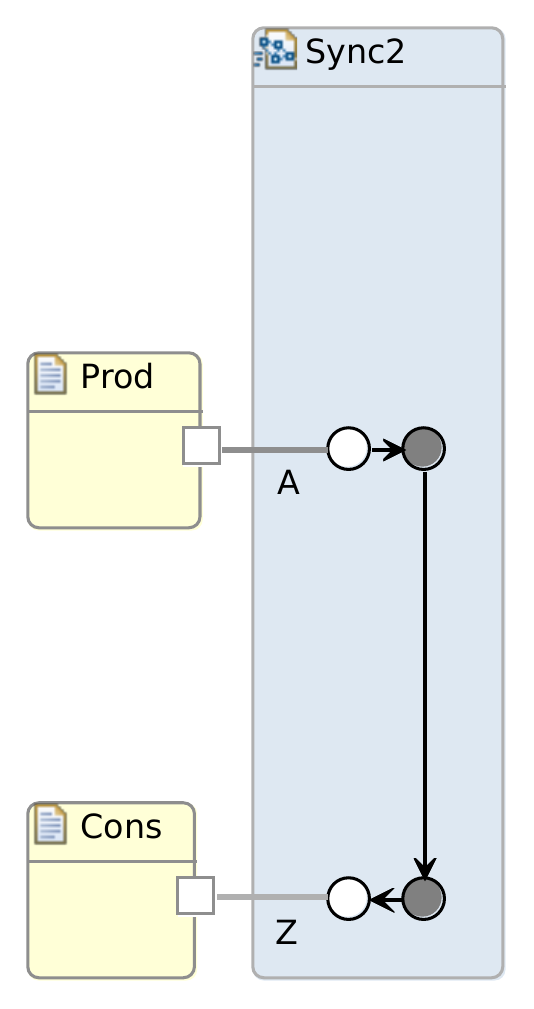}} \vfil
		}
	}
	\hfil
	\subfloat[\reo{Sync3}]{\label{fig:sync:3}%
		\vbox to \HEIGHT {%
			\vfil \hbox {\includegraphics[scale=\SCALE]{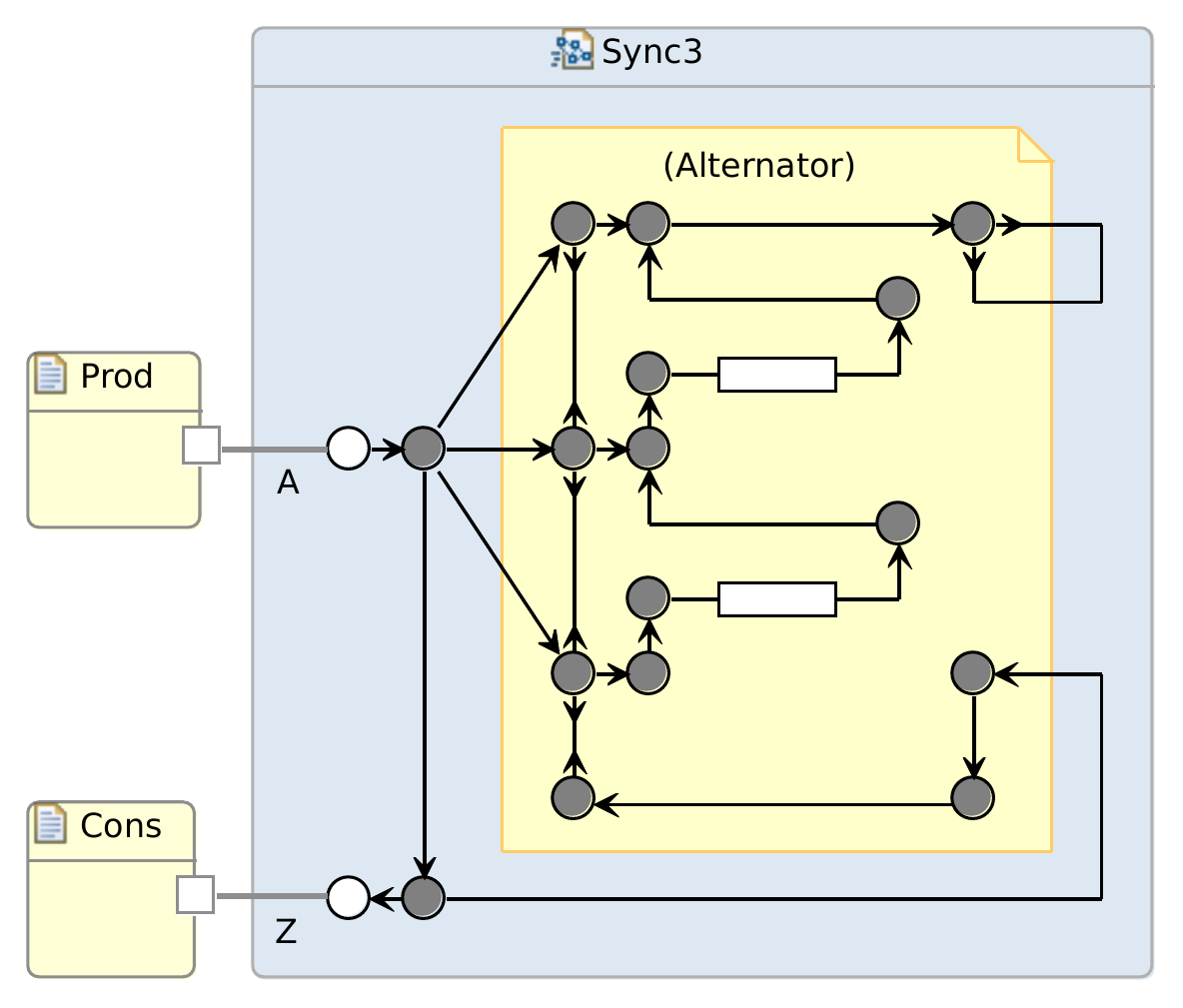}} \vfil
		}
	}
	\hfil
	
	\caption{%
		Behaviorally equivalent connectors, each of which models a standard synchronous channel. 
	}
	\label{fig:sync}
\end{figure}

Generally, two behaviorally equivalent but structurally different connectors may yield different pieces of code with different performance.
Figure~\ref{fig:sync} shows three behaviorally equivalent connectors demonstrating that this applies also to the problems identified in this short paper.
(To see this, note that because the connector in Figure~\ref{fig:sync:3} includes \reo{Alternator}, it suffers from the same problems as \reo{Alternator}).
Consequently, another solution for these problems may be to structurally manipulate connectors (or the sets of small \CA they behave as) before splitting them into regions.
Although we conjecture that such manipulation not always solves our problems, we may identify a class of connectors for which it does.

Finally, at least transition relation explosion may be mitigated by improving our way of dealing with parame\-trization.
In the \reo{Alternator} case, for instance, our current approach to (static) parametrization problematically requires the compiler to compute the \CA for the whole $k$-sized region, given $k$ producers.
A better approach to (static or dynamic) parametrization may enable direct generation of code for $k$ based on the \CA for a 2-sized region without ever computing the \CA for the whole $k$-sized region.

%
\section*{Conclusion}

We introduced two problems---transition relation explosion and overparallelization---with our current compilation approach for Reo.
Intuitively, these problems can be regarded as the flip side of oversequentialization, and its accompanying plague of state space explosion.
Although our first solution works in some cases, a comprehensive solution (including a better understanding of all the cases that this solution should cover), needs to be further developed.
Essentially, we aim at finding the optimal position in the connector implementation spectrum in Figure~\ref{fig:spectrum} that perfectly balances parallelism and sequentiality.

Although encountered by us in the context of Reo, mitigating overparallelization seems a generally interesting problem.
For instance, specifying a system as many parallel processes may feel natural to a system architect, but implementing each of those processes as a thread may give poor performance.
By studying this problem in terms of \CA, which are related to process languages with multiparty synchronization~\cite{KKV12}, we hope to gain new insight and advance compilation technology in areas other than Reo too.
As another example, automatically partitioning \BIP interaction specifications for generating optimal distributed implementations is still an open problem~\cite{BBJ+12,BBQ}.
Further studies may clarify the extent to which the correspondence between \BIP interactions and \CA can be leveraged by reusing results on \CA.

\bibliographystyle{eptcs}
\bibliography{main}

%
\end{document}

%% file: cmds.tex
\newcommand{\cauniv} {\mathbb{C}\mathsc{a}}
\newcommand{\thruniv} {\mathbb{T}\mathsc{hr}}
\newcommand{\thrfun} {\mathsf{thr}}
\newcommand{\prodx} {\boxtimes}
\newcommand{\prodxtt} {\texttt{[x]}}

\newcommand{\BIP}				{\textsc{Bip}\xspace}
\newcommand{\BPMN}				{\textsc{Bpmn}\xspace}
\newcommand{\CA}				{\textsc{ca}\xspace}

\newcommand{\UML}				{\textsc{Uml}\xspace}

\newcommand{\reo}[1]			{\ensuremath{\mathsf{#1}}}

\let\oldsubsubsection\subsubsection
\renewcommand{\subsubsection}[2][]{\oldsubsubsection[#1]{\renewcommand{\,}{\oldcomma} #2}}

\newcommand{\tikzautone}[3]{%
	\ifcompiletikz{%
		\begin{tikzpicture}[baseline]
			\node[state, balanceright, initleft] (Q) {};
			
			\ifthenelse{\equal{#1}{}}{}{%
				\path[->] (Q) edge [trans, out=120, in=60, loop] node [above] {\scriptsize \LINES{#1}} (Q);
			}
			
			\ifthenelse{\equal{#2}{}}{}{%
				\path[->] (Q) edge [trans, out=30, in=-30, loop] node [right] {\scriptsize \LINES{#2}} (Q);
			}
			
			\ifthenelse{\equal{#3}{}}{}{%
				\path[->] (Q) edge [trans, out=-60, in=-120, loop] node [below] {\scriptsize \LINES{#3}} (Q);
			}
			
		\end{tikzpicture}
	} \fi
}

\newcommand{\tikzautonee}[4]{%
	\newcommand{\n}{\protect \\}
	\ifcompiletikz{%
		\begin{tikzpicture}[baseline]
			\node[state, pin={[pin edge={black, thick, <-}]above left:}] (Q) {};
			
			\ifthenelse{\equal{#1}{}}{}{%
				\path[->] (Q) edge [trans, out=105, in=75, loop] node [above] {\scriptsize \LINES{#1}} (Q);
			}
			
			\ifthenelse{\equal{#2}{}}{}{%
				\path[->] (Q) edge [trans, out=15, in=-15, loop] node [right] {\scriptsize \LINES{#2}} (Q);
			}
			
			\ifthenelse{\equal{#3}{}}{}{%
				\path[->] (Q) edge [trans, out=-75, in=-105, loop] node [below] {\scriptsize \LINES{#3}} (Q);
			}
			
			\ifthenelse{\equal{#4}{}}{}{%
				\path[->] (Q) edge [trans, out=-165, in=165, loop] node [left] {\scriptsize \LINES{#4}} (Q);
			}
		\end{tikzpicture}
	} \fi
}

\newcommand{\tikzauttwo}[7]{%
	\newcommand{\n}{\protect \\}
	\ifcompiletikz{%
		\begin{tikzpicture}[baseline, node distance=\tikznodedist, every text node part/.style={align=center}]
			\node[state, initleft] (Q) {};
			\node[state, balanceright, right of=Q] (R) {};
			
			\ifthenelse{\equal{#1}{}}{}{%
				\path[->] (Q) edge [trans, out=120, in=60, loop] node [above] {\scriptsize \LINES{#1}} (Q);
			}
			
			\ifthenelse{\equal{#2}{}}{}{%
				\path[->] (Q) edge [trans, out=30, in=150] node [above] {\scriptsize \LINES{#2}} (R);
			}
			
			\ifthenelse{\equal{#3}{}}{}{%
				\path[->] (R) edge [trans, out=120, in=60, loop] node [above] {\scriptsize \LINES{#3}} (R);
			}
			
			\ifthenelse{\equal{#4}{}}{}{%
				\path[->] (R) edge [trans, out=30, in=-30, loop] node [right] {\scriptsize \LINES{#4}} (R);
			}
			
			\ifthenelse{\equal{#5}{}}{}{%
				\path[->] (R) edge [trans, out=-60, in=-120, loop] node [below] {\scriptsize \LINES{#5}} (R);
			}
			
			\ifthenelse{\equal{#6}{}}{}{%
				\path[->] (R) edge [trans, out=-150, in=-30] node [below] {\scriptsize \LINES{#6}} (Q);
			}
			
			\ifthenelse{\equal{#7}{}}{}{%
				\path[->] (Q) edge [trans, out=-60, in=-120, loop] node [below] {\scriptsize \LINES{#7}} (Q);
			}
			
		\end{tikzpicture}
	} \fi
}